\documentclass[epj]{svjour}
\usepackage{graphics,amssymb,amsmath}
\newcommand{\dphi}{$\Delta\phi$ }
\begin{document}
\title{A Four Component Picture for Jet Induced Correlations in Heavy Ion Collisions?}
\subtitle{Recent Results on Two and Three Particle Correlations}
\author{Anne Sickles% etc
}                     % Do not remove
\institute{Brookhaven National Laboratory, Upton NY 11973}
\abstract{
These proceedings review recent results from two and three particle correlations in
heavy ion collisions.  In particular we discuss the modified structure of the away side
jet correlations.  Under the assumption that the away side can be decomposed into a punch
through component at $\Delta\phi=\pi$ and a shoulder component with a peak displaced
from $\pi$ many similar properties are observed between the ridge and shoulder.  
The particle ratios, yields and $p_T$ spectra are in near agreement.  We also highlight 
important future measurements, including investigating if the decomposition of the away side
jet correlations remains reasonable with high $p_T$ triggers and technical improvements to the
extraction of jet induced correlations.
%
%\PACS{
%      {PACS-key}{describing text of that key}   \and
%      {PACS-key}{describing text of that key}
%     } % end of PACS codes
} %end of abstract
\maketitle
\section{Introduction}
\label{intro}
\label{sec:1}
The main goal of heavy ion collisions is to determine the properties of the hot
dense matter created in central collisions.  While there are many observables of 
interest, jets are very powerful because they are produced only in the initial state.  
The scattered partons then interact with the produced matter from its early stage
until they leave the collision region.  This sensitivity to the time evolution of the 
system and that the initial properties of the jets can be studied in small systems,
p+p and d+Au, provide the ability to study properties of the matter itself and 
how it hadronizes.  

Single particle measurements of jet induced particle have provided constraints
on opacity of the matter~\cite{ppg079}, but multi-particle correlation measurements
provide tomographic information and test the consistency of these models.  Additionally
they are also sensitive to changes in the jet fragmentation process suggested by
single particle results showing a dramatic change in the particle ratios
from small systems to central Au+Au~\cite{ppg015,bmqm08} for $p_T<$5~GeV/c.

Here we describe some of the recent results of jet correlations in heavy ion collisions.

\section{Jet Landscape in Heavy Ion Collisions}
\begin{figure}
\resizebox{\columnwidth}{!}{%
  \includegraphics{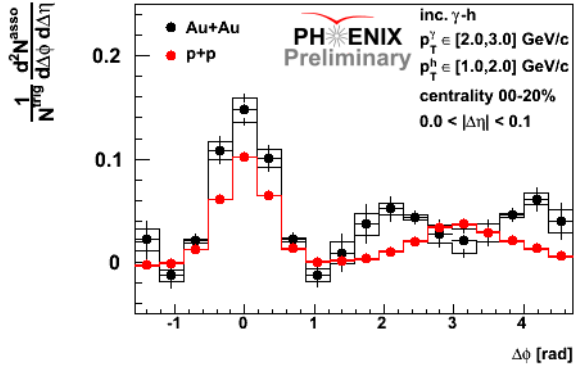}
}
\caption{\dphi distributions of correlated inclusive photon-hadron pairs in central 
Au+Au collisions (black) and p+p collisions (red).  The photons have 2.0$<p_T<$3.0~GeV/c 
and the hadrons have 1.0$<p_T<$2.0GeV/c.  From Ref.~\cite{johnhp08}.}
\label{figdphi_john}       % Give a unique label
\end{figure}

\begin{figure}
\resizebox{\columnwidth}{!}{%
  \includegraphics{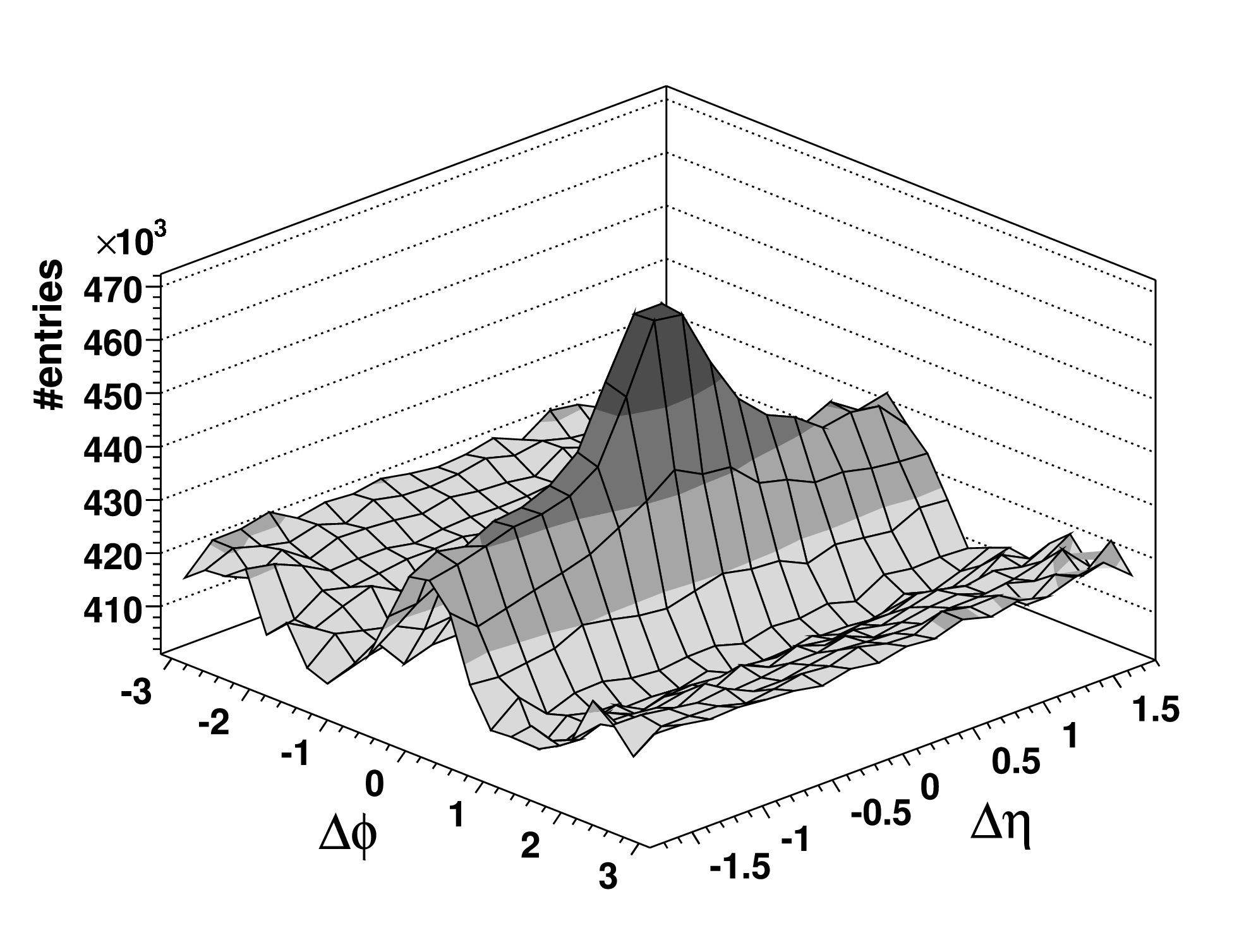}
}
\caption{\dphi v. $\Delta\eta$ distributions of correlated hadron pairs 
in central Au+Au collisions.  Triggers have 3.0$<p_T<$4.0~GeV/c and associated 
particles have 2.0$<p_T<p_{T,trig}$.  From Ref.~\cite{joern_qm06}.}
\label{figridge}      
\end{figure}

The two particle correlation method has been described in detail before (for example,
see Ref.~\cite{ppg083}).
In p+p collisions hard scattering results in two jets approximately back-to-back
in azimuth.
In two-particle
azimuthal correlations this results in an excess of particles at small \dphi and at
$\Delta\phi=\pi$ in what are termed the near and away side jets, respectively.  
These correlations are localized in the $\Delta\eta$ direction as well, however
the di-jets are not constrained to be back to back in $\eta$ because the 
center of mass of the parton-parton scattering is not necessarily the collision
center of mass.  
One of the most striking results from RHIC is that this qualitative picture is dramatically
changed in central collisions at RHIC; two new structures are observed.  There is a peak
on the away side displaced from $\Delta\phi=\pi$ by approximately one radian, 
the {\it shoulder} (see
Fig.~\ref{figdphi_john}) and 
an elongated structure in $\Delta\eta$ on the near side, the 
{\it ridge} (see Fig.~\ref{figridge}).   The origins and properties of these structures
are under investigation and for the purposes of these proceedings we will treat the
observed jet correlations in this four component picture: the near side jet (small
$\Delta\phi$, small $\Delta\eta$); the ridge (small $\Delta\phi$, wide in $\Delta\eta$);
the away side punch through jet ($\Delta\phi\thickapprox\pi$); and the shoulder 
($\Delta\phi\thickapprox\pi\pm$1).  It is still a matter of experimental investigation
to what extent the shoulder and punch through distinction is the right picture.
However, by making this separation and looking at the properties of the extracted
shoulder suggestive similarities are seen to the properties of the ridge.

\subsection{The Ridge}
\paragraph{Extraction of the Ridge}
The ridge yield is extracted by measuring small $\Delta\phi$ correlations in
a $\Delta\eta$ region that is expected to have only a small yield from jets.
Jana Bielcikova spoke at this conference~\cite{janahp08} about 
what is known experimentally about the ridge
in heavy ion collisions as well as theoretical ideas about its origin.  The ridge is
extended in $\Delta\eta$; currently there are no measurements of its full
extent; PHOBOS
sees evidence for a ridge out to $\Delta\eta$=4~\cite{wengerqm08}, the end of their acceptance
in correlations between trigger hadrons with $p_T>$2.5GeV/c and partner hadrons
with $p_T>$20MeV/c.
%This, in itself, provides constraints on 
%what the source of the ridge can be because of the need to obey causality; late stage effects
%could simply not propagate information over 4 units in $\Delta\eta$~\cite{raju}.

\subsection{The Shoulder}

\begin{figure}
\resizebox{\columnwidth}{!}{%
  \includegraphics{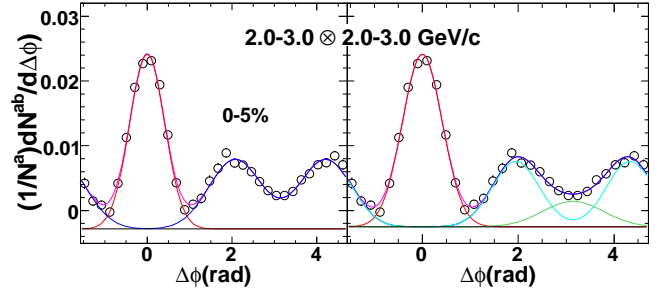}
}
\caption{Fits to distributions of correlated hadrons.  The left panel shows the away
side described by a two Gaussian fit and the right panel shows the away side described
by a three Gaussian fit with the additional Gaussian centered at $\Delta\phi=\pi$.  
Both fits describe the data relatively well.  From Ref.~\cite{ppg083}.}
\label{figdphi}      
\end{figure}

\paragraph{Extraction of the Shoulder}
\begin{figure}
\resizebox{\columnwidth}{!}{%
  \includegraphics{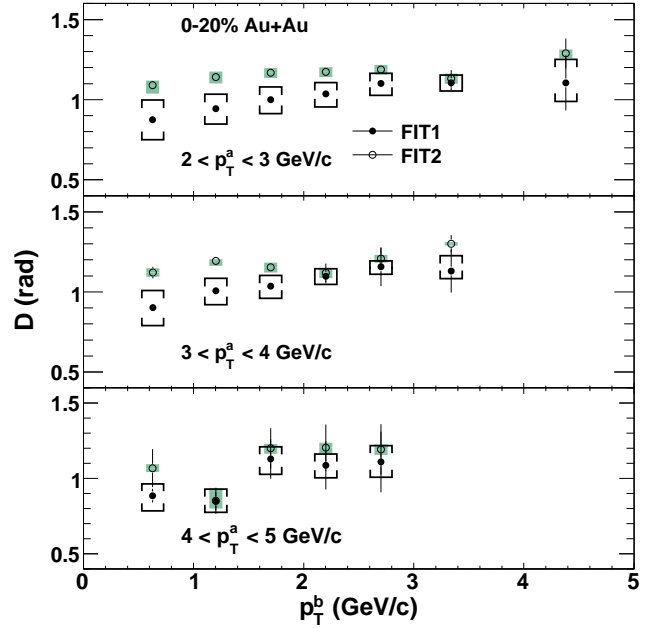}
}
\caption{$D$ extracted with two-Gaussian (solid points) and three-Gaussian (hollow points)
 fit from correlations between hadron
pairs in the $p_T$ selections indicated on the plot.  From Ref.~\cite{ppg083}.}
\label{dppg083}      
\end{figure}

\begin{figure}
\resizebox{\columnwidth}{!}{%
  \includegraphics{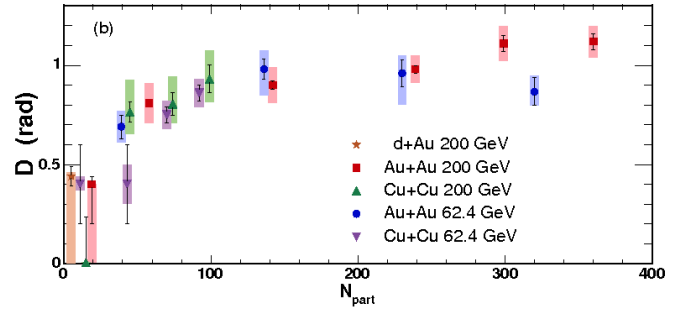}
}
\caption{$D$ extracted with a two-Gaussian fit from correlations between hadron
pairs with triggers at 2.5$<p_T<$4.0GeV/c and associated particles with 1.0$<p_T<$2.5GeV/c.  
From Ref.~\cite{ppg067}.}
\label{dppg067}      
\end{figure}

\begin{figure*}
\resizebox{\textwidth}{!}{%
  \includegraphics{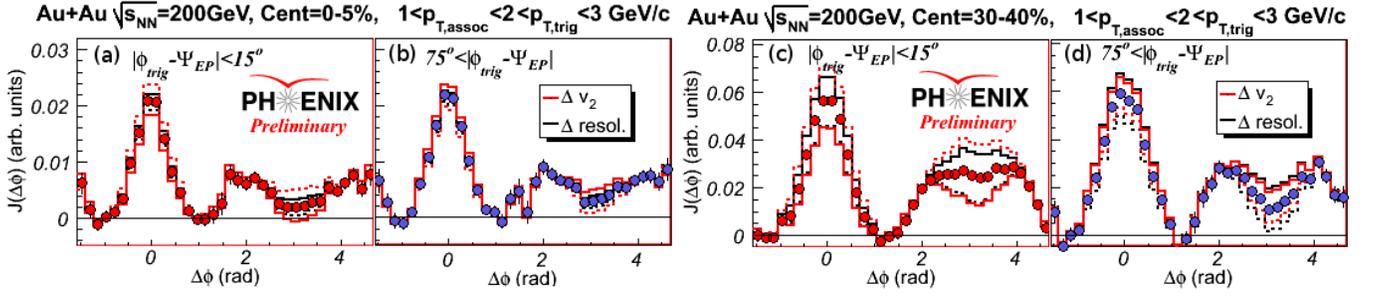}
}
\caption{Background subtracted distributions of hadrons with 1.0$<p_T<$2.0 associated
 with trigger hadrons with 2.0$<p_T<$3.0GeV/c where the trigger has been selected
to be in the reaction plane (panels (a) and (c)) or out of the reaction plane (panels (b)
and (d)).  Panels (a) and (b) are 0-5\% central collisions and (c) and (d) are mid-central
(30-40\%) collisions. From Ref.~\cite{mikeqm08}.}
\label{rxnp_phenix}      
\end{figure*}

\begin{figure*}
\resizebox{\textwidth}{!}{%
  \includegraphics{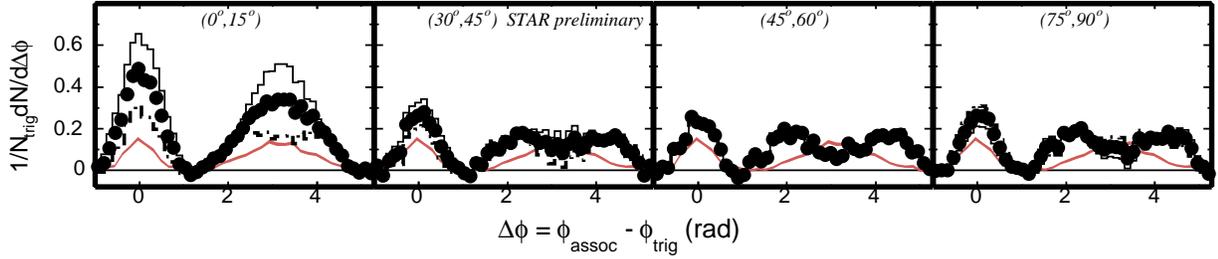}
}
\caption{Background subtracted distributions of hadrons with 1.0$<p_T<$1.5GeV/c associated
with trigger hadrons with 3.0$<p_T<$4.0GeV/c where the trigger orientation
with respect to the reaction plane has been chosen to be within the angular window indicated
on the panel.  The Au+Au events have centrality from 20-60\%.  From Ref.~\cite{marcohp08}.}
\label{rxnp_star}      
\end{figure*}

The PHENIX Collaboration has used a variety of techniques to extract information about the shoulder
from two-particle correlations. Fig.~\ref{figdphi}~\cite{ppg083}
 shows two different methods to extract the shoulder
location under the assumption of a Gaussian shape.  On the left is a two Gaussian fit with the 
away side Gaussians displaced from $\pi$ by $\pm D$~rad.  On the right is a three-Gaussian fit which
includes an additional Gaussian at $\Delta\phi=\pi$ corresponding to punch-through jets,
those back-to-back correlations as in p+p.  With both two- and three-Gaussian fits PHENIX
has found that the $D$ parameter increases slightly with  $p_T$ for central collisions, 
Fig~\ref{dppg083}.
Interestingly, no significant dependence on either collision
energy or the colliding nuclei is seen either.  Fig~\ref{dppg067}  shows $D$ 
for Au+Au and Cu+Cu collisions
at both 200 and 62.4GeV center of mass energy extracted from a two-Gaussian fit.  
These results pose a challenge to identifying the shoulder structure
with a Mach Cone shock-wave since the angle of the shoulder should be~\cite{jorge1105}:
$\cos\theta_M = \frac{\bar{c_s}}{v_{jet}}$
where $v_{jet}$ is the velocity of the jet (essentially 1 for light quark jets)
and $\bar{c_s}$ is the time averaged speed of
sound in the matter, which change with the initial temperature and lifetime of the system.

Three particle correlations are sensitive to correlations between particles
in the shoulder.    Three particle
correlation results from STAR, what is observed as a shoulder in two particle correlations 
is actually consistent with 
the projection of a cone structure on the away side~\cite{star3part}.  However the 
opening angle observed
in the three particle correlation measurements 
($\theta$=1.38 $\pm$ 0.02(stat) $\pm$ 0.06(sys)~\cite{star3part}
is significantly larger that the distance between the peaks seen in two particle 
correlations in a similar $p_T$ range~\cite{ppg083,ppg067}.  
CERES also has three particle correlation results that show evidence of conical
structure on the away side~\cite{kniegeqm06}, but a cone angle has not
been extracted.

\paragraph{Reaction Plane Dependence}
Figs.~\ref{rxnp_phenix} and~\ref{rxnp_star} show hadron-hadron correlations 
where the trigger particle is also selected based on it's angle with respect to the
reaction plane.  For the central collisions, Fig.~\ref{rxnp_phenix} panels (a) and (b),
there is no difference whether the trigger is in or out of the reaction plane.  This
is expected because when the impact parameter is small the initial state is fairly
symmetric.  In the mid-central collisions, Fig.~\ref{rxnp_phenix} panels (c) and (d)
and Fig.~\ref{rxnp_star}, there is a greater variation in the central values, going
from a single peak away side structure when the trigger is in plane to a double
peaked away side structure when the trigger is out of plane.  This is qualitatively
consistent with a greater suppression of the punch through component of the away side
jet when the trigger particle is out of plane as would be expected from the away side
parton having a longer matter path length.  However, the systematic errors from the
$v_2$ values used in the background subtraction must be considered.  As can
be seen in Fig.~\ref{rxnp_phenix} the effects of the $v_2$ error are anti-correlated
between in-plane and out of plane triggers.  Pushing these measurements to higher $p_T$
where the combinatoric background levels are smaller will help constrain these shapes.
In all cases the yield in the shoulder region is nearly constant with respect to the 
trigger orientation with respect to the reaction plane.

\subsection{Connections Between the Ridge and the Shoulder}

\paragraph{Particle Ratios}

\begin{figure}[h]
\resizebox{\columnwidth}{!}{%
  \includegraphics{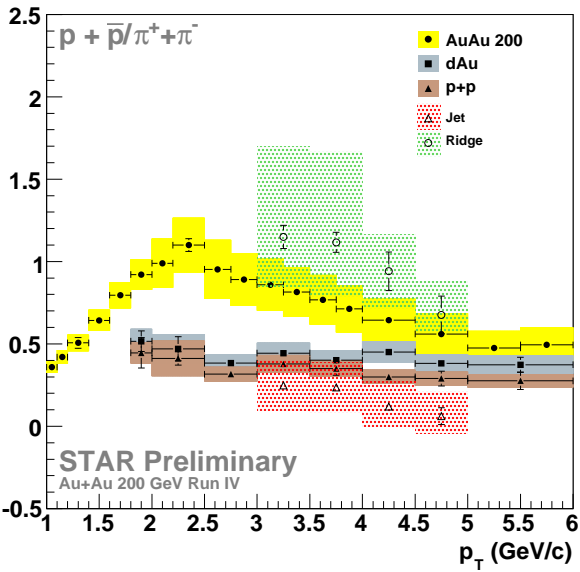}
}
\caption{Ratio of $(p+\bar{p})/(\pi^+ + \pi^-)$ for the ridge, near side jets for 
4$<p_{T,trig}<$6GeV/c.  Inclusive particle ratios from Au+Au, d+Au and p+p collisions 
are shown for comparison.  From Ref.~\cite{marcohp08}.}
\label{ridge_ratio}      
\end{figure}

Fig.~\ref{ridge_ratio} from STAR~\cite{marcohp08} shows the particle composition 
of the ridge to have a high
$\frac{p+\bar{p}}{\pi^{\pm}}$ ratio, similar to Au+Au inclusive particle ratios 
and greater than in the near side jet, where the inclusive particle 
ratio is similar to p+p and d+Au collisions.  
%These results could suggest that particle
%production in the ridge occurs in a manner similar to the inclusive particles.
%At 2$<p_T<5$GeV/c, inclusive particle production is thought to be dominated by 
%recombination and that perhaps explains the ridge~\cite{hwaqm08}.

\begin{figure}
\resizebox{\columnwidth}{!}{%
  \includegraphics{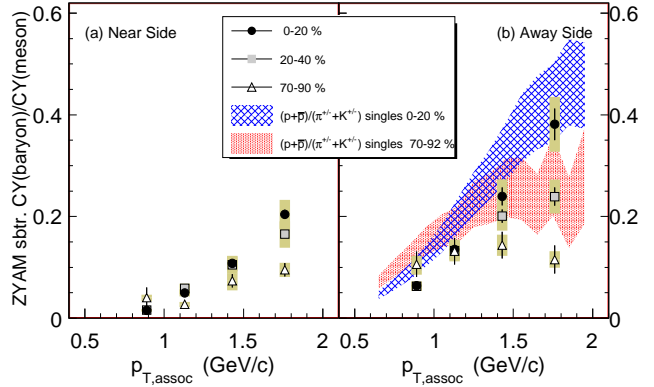}
}
\caption{Ratio of $(p+\bar{p})/(\pi^{\pm} + K^{\pm})$ for particles associated with a hadron trigger at
2.5$<p_T<$4.0GeV/c.  The left panel shows the ratio for the near side correlations (the
ridge has not been subtracted  and the right for the entire away side region including both
the shoulder and punch-through jet region.  The shaded regions show the same ratios extracted
from single particle spectra in the centrality ranges shown.
From Ref.~\cite{ppg034}. }
\label{ppg034_fig4}      
\end{figure}

PHENIX has studied the particle ratios for particles associated with a hadron trigger
over the entire away side $\Delta\phi$ range and finds the ratio of $\frac{p+\bar{p}}
{\pi^{\pm} + K^{\pm}}$ to be significantly enhanced with increasing collision 
centrality (see Fig~\ref{ppg034_fig4})~\cite{ppg034}.  In the most central collisions,
the ratios of associated particles are in agreement with those for inclusive particles,
suggesting that the same hadronization mechanism could be responsible for the particles in the 
ridge, away side and inclusive particles.  It has been speculated that as the fast
parton traverses the produced matter and loses energy localized regions of
increased energy density are produced~\cite{friescs,rudyqm08,ppg034}.  
When these regions hadronize they produce
particles that are correlated with the hard scattering, but have properties
more similar to that of the bulk.

However, care should be taken in the interpretation of the current results.
Extraction of the ridge and shoulder associated particle ratios within the same analysis
is important.  While both the ridge and shoulder ratios in Figs.~\ref{ridge_ratio} 
and~\ref{ppg034_fig4} appear consistent with the inclusive ratio reference, the associated particle
ratios themselves cannot be compared directly.  The PHENIX measurement includes $K^{\pm}$
in the denominator and is for both the shoulder and the away side jet region.  
Additionally the centrality selections and the trigger hadron $p_T$ ranges are different.
Eliminating these differences will allow for a clean comparison.

\begin{figure}
\resizebox{\columnwidth}{!}{%
  \includegraphics{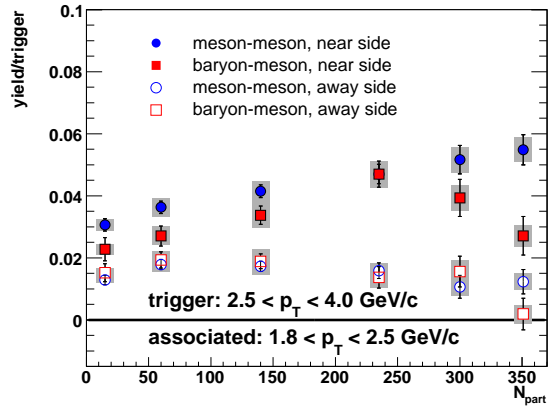}
}
\caption{Yield per trigger on the near, $\Delta\phi<0.94$rad (solid points) and 
away 
$\Delta\phi>$2.2rad (hollow
points) side for baryon-meson (squares) and meson-meson (circles) 
correlations as a function of $N_{part}$.  Triggers have 2.5$<p_T<$4.0~GeV/c
and associated particles have 1.8$<p_T<$2.5~GeV/c.
Error bars are statistical
errors and the shaded boxes show the systematic errors.  There is a
13.6\% normalization error which moves all points together. 
From Ref.~\cite{ppg072}. }
\label{ppg072_fig7}      
\end{figure}

Fig.~\ref{ppg072_fig7}~\cite{ppg072} shows yields from correlations where both particles
have been identified as baryons or mesons.  Both particles are 
inside the PHENIX $\eta$ acceptance $|\eta|<$0.35 and the ridge and jet
components have not be separated.  The meson-meson near side yields
increase with centrality, qualitatively consistent with an increasing
ridge component in addition to the same side jet.  The near side baryon-meson yields
also increase with centrality up to $N_{part}\approx$250, but then decrease
for the most central collisions.  This might occur if the baryons
originate dominately from the ridge and the near side jet correlation is reduced,
possibly because it is outside the experimental $\Delta\eta$ acceptance.
Interestingly, for the away side (here
with $|\pi-\Delta\phi|<$0.94~rad) no dependence is seen on whether the trigger
particle is a baryon or a meson.   

\begin{figure}
\resizebox{\columnwidth}{!}{%
  \includegraphics{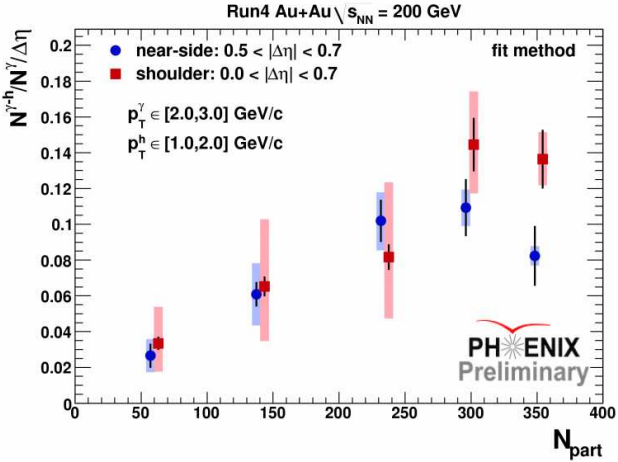}
}
\caption{Yield of associated hadrons per trigger per $\Delta\eta$ for both the ridge
and the shoulder as a function of $N_{part}$~\cite{johnhp08}.  The triggers are inclusive photons
mainly from $\pi^0$ decay with 2.0$<p_T<$3.0GeV/c
and the associated particles are hadrons with 1.0$<p_T<$2.0GeV/c.}
\label{john_yields}      
\end{figure}

\paragraph{Centrality Dependence}
PHENIX using a three-Gaussian fitting method and inclusive photon trigger (which
at moderate $p_T$ primarily come from $\pi^0$ decay where the trigger photon
carries most of the $\pi^0$ $p_T$) has extracted the yield per trigger,
per $\Delta\eta$ for both the ridge and the shoulder, Fig~\ref{john_yields}.
The yields for both the ridge and the shoulder increase from peripheral to central collisions.
The magnitudes of the yields are consistent at all centralities except perhaps the 
most central collisions.

\paragraph{$p_T$ Spectra}

\begin{figure}
\resizebox{\columnwidth}{!}{%
  \includegraphics{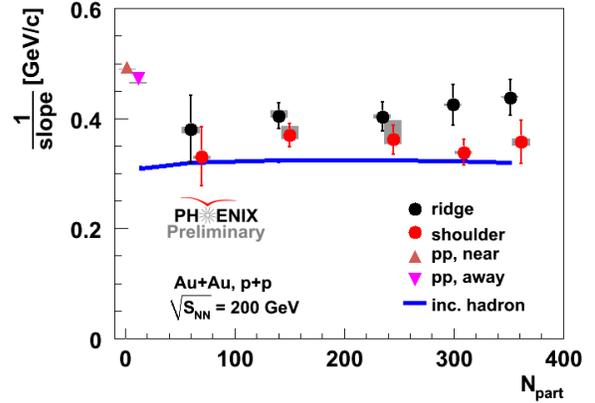}
}
\caption{Inverse slopes extracted from exponential fits to the $p_T$ spectra of hadrons with
1.0$<p_T<$5GeV/c associated with inclusive photons with 2.0$<p_T<$3.0GeV/c for ridge and shoulder
correlations.  For comparison near and away side p+p results for the same $p_T$ ranges are shown
along with the inclusive hadron inverse slopes.  From Ref.~\cite{johnhp08}.}
\label{john_slopes}      
\end{figure}

Also within the same fitting analysis, Fig.~\ref{john_slopes} 
shows the slope of the associated $p_T$ spectra as a function
of $N_{part}$.  Here again, within the experimental uncertainties the ridge and spectra slopes
are similar.  The ridge spectra are slightly harder than the shoulder spectra and both
are significantly softer than jets in p+p collisions.

Experimentally, reduced error bars, available 
from the data already on tape, and more measurements (for example, Cu+Cu collisions, larger $p_T$ ranges,
etc) are needed to clarify the situation.  Theoretically, most models only address the ridge
or the shoulder; models which could possibly explain the connection between them could help
clarify the relationship between the ridge and the shoulder.  
%Renk and Ruppert have speculated the ridge is in fact an incompletely
%formed shock wave and the more completely formed shock waves are seen as the shoulder on the
%away side~\cite{renk_hangzhou}.  
%Additionally, the ridge could possibly result from triggering
%on particles that are part of the matter's response to the passage of a fast parton.  Thus,
%it could be worthwhile to investigate whether the ridge and a near side jet correlation are seen in 
%the same event.

The preliminary connections between the ridge and the shoulder are very interesting.  Certainly 
more experimental studies are needed.  To eliminate ambiguities both the ridge and the shoulder
should be studied in the same analysis.   The large Au+Au
statistics from the 2007 data set should be very powerful however given the observed
away side in smaller systems and at lower energy, studies of the ridge in such systems is also
important.

\section{Outlook}
\subsection{Jets at High $p_T$}

At high $p_T$ jet measurements are currently limited by statistics.
Here it is less clear whether the shoulder/punch through jet distinction used above
is still meaningful or whether the away side is best described as a single, perhaps
widened,  peak centered at $\Delta\phi=\pi$.  
In order to answer this question the trigger should have
a high $p_T$ (greater than 7GeV/c) and be identified as a pion or direct photon
 to eliminate ambiguities
arising from baryons.  Initial work has been done 
in both STAR~\cite{hamedhp08} and  PHENIX~\cite{andrewqm08,justinhp08}.  
At intermediate $p_T$ the extraction of $D$ depends
little on the fitting method used (see Fig~\ref{dppg083}) since the shoulder peak is 
visible and the yield at $\Delta\phi=\pi$ is suppressed.  However, at higher trigger $p_T$
the away side peak appears as a single broad peak or a flat distribution around $\Delta\phi=\pi$,
so the conclusions will be more sensitive to how the parameters are extracted.

%\paragraph{Is There a Punch Through + Shoulder or a Widened Jet?}
%If the shoulder structure is due to Mach Cones it should be independent of trigger 
%$p_T$~\cite{pjwwnd08}.  Fig.~\ref{ppg083_trigpt} shows initial 
%results with correlations between two charged hadrons 
%show a weak dependence of the yield in the shoulder (extracted within the fixed angular range
%$\pi/6 < |\Delta\phi| < \pi/3$) on the trigger hadron $p_T$.  When both trigger and partner $p_T$s
%are large the yield in the shoulder region increases more quickly with trigger particle
%$p_T$, though this could perhaps be due to larger punch through component leaking in
%from the head region.  Future measurements with a high $p_T$ identified particle trigger,
%smaller trigger $p_T$ ranges, more sophisticated shoulder extraction and of course better
%statistics will be crucial to understanding this issue.  Work is ongoing~\cite{andrewqm08}.

%\begin{figure}
%\resizebox{\columnwidth}{!}{%
%  \includegraphics{shoulder_yield_vtrig_ppg083.eps}
%}
%\caption{Data from Ref.~\cite{ppg083}. }
%\label{ppg083_trigpt}      
%\end{figure}

%\paragraph{Geometrical Bias, 2+1 Correlations}

%\section{Precision Measurements}
%\subsection{Particle Identification at Intermediate $p_T$}

%\subsection{Heavy Flavor and Photon Correlations}
%\subsection{Fragmentation Photons}
\subsection{Absolute Subtraction Method}
The results shown here were with the exception of those from Ref.~\cite{ppg072} have
the combinatoric background subtraction under the assumption that there is a region
in $\Delta\phi$ which has no signal (zero yield at minimum--ZYAM).  
However, the signal distributions are wide
in comparison with their distance apart so this ZYAM assumption must result in a 
lower limit on the jet signal.  An alternative method of extracting the jet functions,
the absolute subtraction method~\cite{ppg072,anne_hangzhou} makes no assumption
about the shape of the signal.  In the case of moderate statistics the two background
subtraction methods have been shown to agree~\cite{ppg067}.  
However, as seen in Fig.~\ref{figdphi} with good statistical precision there can
be a discrepancy about the shape parameters extracted from the jet functions
and the yields extracted under the ZYAM assumption.  Also, in the case of poor
statistics the absolute background subtraction method provides an improvement
over ZYAM because large statistical errors in the correlation functions
propagate into large errors on the background level, which lead to large errors on
the extracted parameters.  

\section{Conclusion}
We have focused on correlations between two hadrons in Au+Au collisions at 200GeV.
The jet functions in central heavy ion collisions are qualitatively different than
in peripheral heavy ion or p+p collisions with the addition of the ridge and shoulder
structures.  Initial results show that the yields, particle ratios, and spectra
shape are similar between these two structures.  This could indicate that the source
of the shoulder and the ridge is similar or that they are sensitive to similar properties
of the matter produced in heavy ion collisions.  However, further experimental work
with high $p_T$ triggers and selection of the trigger with respect to the reaction plane
needs to be done to determine if the separation of the away side into the shoulder
and punch through components is justified.
 Additionally, results from two and three particle 
particle correlations at lower energies show a shoulder structure similar to that
seen at 200GeV.  This challenges the attribution of the shoulder at 200GeV to 
a Mach Cone shock-wave.  We look forward to further results incorporating 
more sensitive observables, more particle identification,   and more statistics.

% BibTeX users please use
 \bibliographystyle{iopart-num.bst}
 \bibliography{sickles_hp08}

\end{document}